\documentstyle[preprint,aps,prl,floats,epsf]{revtex}  
  
\newif\ifmad
\madfalse 

\ifmad
\makeatletter
\ifx\every@math@size\undefined\else  
  \let\old@expast\@expast  
  \def\@expast#1{\old@expast{#1}\let\@tempa\reserved@a}  
\fi  
\makeatother  
\fi

\tightenlines 
  
\preprint{  
\hbox to \hsize{  
\hfill$\vcenter{\hbox{\bf MADPH-01-1231}  
	        \hbox{\bf OKHEP-01-04}  
                \hbox{\bf hep-ph/0106189}  
                \hbox{June 2001}}$ }  
}  
  
\begin{document}  
  
\title{\vspace*{.25in} 
Implications of new CMB data for neutralino dark matter} 
  
\author{V. Barger$^a$ and Chung Kao$^b$}  
  
\address{  
$^a$Department of Physics, University of Wisconsin,  
Madison, WI 53706 \\  
$^b$Department of Physics and Astronomy, University of Oklahoma,  
Norman,  OK 73019}  
  
  
\maketitle  
  
\thispagestyle{empty}  
  
\begin{abstract}  
 
The combination of new cosmic microwave background data with  
cosmological priors has determined the physical cold dark matter (cdm)  
density to be $\Omega_{\rm cdm} h^2 = 0.13\pm0.01$. We find the  
corresponding regions of parameters in the minimal supergravity  
model for which the lightest neutralino is the cdm particle.   
We then compare with the muon anomalous magnetic moment ($a_\mu$)  
measurement, the mass of the lighter CP-even Higgs boson ($h^0$), 
the trilepton search at the upgraded Fermilab Tevatron collider,   
and a direct search for relic neutralinos.   
The intersection of $\Omega_{\rm cdm} h^2$ and $a_\mu$ constraints 
selects $\tan\beta\approx 40-45$.  
 
\end{abstract}

\newpage  
  
Recent precision measurements of the Cosmic Microwave Background (CMB)     
anisotropy~\cite{Boomerang,DASI,Maxima} place restrictive constraints   
on the densities  of matter and dark energy in the Universe.   
The results of parameter extraction from the CMB data made in conjunction with 
other cosmological priors (flat universe, supernovae, large-scale structure) 
give a baryon density\footnote{The Hubble constant at the present time 
is expressed as $H_0 =  100 h \; \rm km\; s^{-1} \; Mpc^{-1}$.}$^,   
$\footnote{The relative density, $\Omega = \rho/\rho_{\rm c}$,  
is the ratio of the density $\rho$ to the critical density  
$\rho_{\rm c}$ which would close the Universe.}  
$\Omega_b h^2  = 0.022\pm0.003$, which is very consistent with the  
predictions of Big Bang Nucleosynthesis theory.   
A cold dark matter density $\Omega_{\rm cdm} h^2 =  0.13\pm0.01$   
is extracted \cite{Boomerang,DASI}. 
This restrictive range for $\Omega_{\rm cdm} h^2$  
has significant impact on the allowed masses of cold dark matter    
particles of supersymmetry origin, which we study here.  
   
The lightest supersymmetric particle (LSP) of a supersymmetry theory  
with a conserved $R$-parity\footnote{The Standard Model particles and  
Higgs bosons have $R=+1$ and their superpartners have $R=-1$.}  
is a preferred candidate for cold dark matter. 
In the minimal supergravity (mSUGRA) model \cite{sugra}, 
supersymmetry (SUSY) is broken in a hidden sector and SUSY breaking 
is communicated to the observable sector through gravitational interactions.  
For most of the mSUGRA parameter space, the LSP is the lightest neutralino  
($\chi_1^0$), which is a linear combination of gauginos  
($\widetilde B, \widetilde W_3$) and and higgsinos ($\widetilde    
H_1, \widetilde H_2$) that are superpartners of gauge bosons ($B, W_3$)  
and Higgs bosons ($H_1, H_2$).  
  
Neutralinos existed abundantly in the early Universe in thermal  
equilibrium with other particles; their pair annihilations were balanced  
by pair creation. As the Universe cooled, the neutralino density became  
Boltzmann-suppressed.   
Deviation from thermal equilibrium began when the temperature reached  
the freeze-out temperature  $T_f \simeq m_{\chi^0_1}/20$.  
After the temperature dropped to $\sim{1\over5}T_f$,  
the annihilation rate became equal to the expansion rate and  
$n_{\chi^0_1} = H/\left<\sigma v\right>$, where $H$ is the Hubble  
expansion rate at that temperature.  
Here $\left<\sigma v\right>$ is the thermally averaged cross section  
times neutralino velocity\cite{jkg}.   
The relic mass density became 
\begin{equation}  
\Omega_{\chi_1^0} h^2 = n_{\chi_1^0} m_{\chi_1^0} / \rho_{\rm c}  = H    
m_{\chi_1^0} / \left( \left< \sigma v \right> \rho_c \right)  
\end{equation}  
and the neutralinos remained as cold dark matter.    
  
\bigskip\noindent  
\underline{\bf mSUGRA Model\ }  
The predicted $\Omega_{\chi_1^0}$ in the mSUGRA model increases with    
$m_{\chi_1^0}$, except where resonances of neutral Higgs bosons ($H^0,A^0$) 
enhance $\left<\sigma v\right>$ and thereby suppress $\Omega_{\chi_1^0}$. 
Previous analyses made to limit the mSUGRA mass parameters allowed broad  
regions of the cdm density, typically  
$0.1 \alt\Omega_{\chi_1^0} h^2 \alt 0.5$~\cite{run2}.  
With the very restrictive range now selected from the CMB data and priors, 
we can expect much tighter constraints on the mass spectrum 
of supersymmetric (SUSY) particles. 
In the present study we consider the $2\sigma$ cdm density range    
$0.11\alt \Omega_{\chi_1^0} h^2 \alt 0.15$ determined by the CMB analyses.  
We evolve supersymmetry masses and couplings from the grand unified  
scale using two-loop renormalization equations~\cite{run2}.  
The mSUGRA parameters are a scalar mass ($m_0$),  
a gaugino mass ($m_{1/2}$), a trilepton coupling ($A_0$),  
the sign of a Higgs mixing parameter ($\mu$), and  
the ratio of Higgs field vacuum expectation values at the electroweak  
scale ($\tan\beta = v_2/v_1$).   
The value of $A_0$ only significantly affects results for high $\tan\beta$; 
we initially take $A_0 = 0$ and study the $A_0$ dependence later.
Moreover, we focus on $\mu>0$, in the sign convention of Ref.~\cite{run2}, 
since positive $\mu$ is preferred by mSUGRA  
calculations of the $b\to s\gamma$ decay branching  
fraction~\cite{bsg1,bsg2} and required if the $2.6\sigma$   
deviations of the muon anomalous magnetic moment  
$a_\mu={1\over2}(g-2)_\mu$ from the Standard Model (SM) \cite{muon1}   
are of supersymmetry origin\cite{muon2,drees,baer01}.  
  
\bigskip\noindent  
\underline{\bf CDM Density\ }  
The dark blue bands in Figure 1 show regions 
in the $m_0$ versus $m_{1/2}$ plane 
where the predicted $\Omega_{\rm cdm} h^2$ is within the CMB $2\sigma$ range.  
The four panels of the figure display $\tan\beta$ values of 3, 10, 35, and 50. 
The chargino search at LEP\,2 \cite{SUSY} excludes 
$m_{\chi_1^\pm} < 103$~GeV, 
and the Higgs search at LEP\,2 excludes $0.5<\tan\beta < 2.4$ \cite{Higgs}.
At $\tan\beta = 3$, 10 and 35, there are pockets of allowed masses 
with $m_0,m_{1/2}\alt 400$~GeV. We note that the lightest neutralino mass 
is related to $m_{1/2}$ by \cite{omhs97}  
\begin{equation}  
m_{\chi_1^0} \simeq 0.435m_{1/2} -2.8 \sin2\beta -10.4 
\end{equation}  
and the lightest chargino mass is given by  
$m_{\chi_1^\pm} \approx 2m_{\chi_1^0}$.  
The low mass pockets at $\tan\beta = 3$ and 10 allowed by the CMB range of  
$\Omega_{\rm cdm} h^2$ have chargino and neutralino masses that are  
accessible at Run~II of the Tevatron collider\cite{run2}.   
There is an additional narrow allowed band extending to high $m_{1/2}$,   
where $\chi^0_1 - \tilde\tau_1$ coannihilation dominates\footnote{  
We use the calculations of coannihilation in Ref.~\cite{Toby}.}   
\cite{Toby,Arnowitt:2001yh};   
the region just below this band is excluded because the charged  
$\tilde\tau_1$ would be the LSP there.  
When $\tan\beta$ is increased to 35, the allowed masses in the pocket region 
increase and also a new narrow window opens at high $m_0$, just below  
the theoretically excluded boundary from electroweak symmetry breaking.   
Then at $\tan\beta=50$, the approximate maximum value for which the  
theory is perturbative, solutions exist only at high $m_0$ and $m_{1/2}$  
masses. The allowed bands and the empty regions in Fig.~1(d)  
can be qualitatively understood in terms of the $s$-channel  
Higgs resonance effects.  
  
\bigskip\noindent  
\underline{\bf Muon $\bf g-2$\ }  
The reported $2.6\sigma$ deviation of the muon anomalous magnetic  
moment~\cite{muon1} from its predicted SM value has triggered  
supersymmetry explanations (see e.g.\ Ref.~\cite{muon2,baer01}).   
Interestingly, the low mass pockets in Figs.~1(a,b,c) from our cdm  
analysis have considerable overlap with the $2\sigma$ region favored  
by the E821 measurement~\cite{baer01}. The dashed curves in Fig.~1  
show the mSUGRA predictions for   
\begin{eqnarray}  
\Delta a_\mu = \left[a_\mu - a_\mu({\rm SM})\right] /  10^{-10}. 
\label{eq:damuon}  
\end{eqnarray}  
The curves represent 
the central value of the measurement $\Delta a_\mu(\rm exp)=43$ and   
its $2\sigma$ uncertainties $\Delta a_\mu({\rm exp}) \pm  2\sigma = 75, 11$.  
At large $\tan\beta$, the chargino-sneutrino loop diagram 
is the dominant SUSY contributor to $\Delta a_\mu$.  
  
\bigskip\noindent  
\underline{\bf Lightest Higgs Boson\ }  
The LEP\,2 collaborations found a tentative signal for a Higgs boson  
with mass of 115~GeV \cite{Higgs}. The contours of mSUGRA parameters  
for which this mass is predicted are shown in Figure~1.  
We calculate the masses and couplings in the Higgs sector with one-loop  
corrections from the top, the bottom and the tau Yukawa interactions  
in the RGE-improved one-loop effective potential \cite{one-loop}  
at the scale $Q = \sqrt{m_{\tilde{t}_L}m_{\tilde{t}_R}}$ \cite{Baer97}.  
With this scale choice, the RGE improved one-loop corrections  
approximately reproduce the dominant two-loop perturbative calculation  
of the mass of the lighter CP-even Higgs scalar ($m_h$) \cite{two-loop}.  
We note that pocket regions at $\tan\beta = 3, 10$ have 
$m_h \alt 100$ GeV, 115 GeV respectively, 
and are thus not favored by the LEP\,2 Higgs search. 
However, at $\tan\beta = 10$, the coannihilation band overlaps 
with $m_h \agt 115$ GeV.

\bigskip\noindent  
\underline{\bf Trileptons at Run~II\ }  
Systematic studies have been made of potential for supersymmetry  
discovery in Run~II at the upgraded Fermilab Tevatron collider  
\cite{run2}. The most promising signal is trileptons from  
chargino--neutralino ($\chi_1^\pm \chi_2^0$) associated production.  
Figure~1 shows the $3\sigma$ signal  contours with 30~fb$^{-1}$  
luminosity \cite{trilepton}. 
The CMB allowed mass regions at $\tan\beta \alt 10$ have considerable  
overlap with the trilepton signal regions accessible at Run~II.  
  
\bigskip\noindent  
\underline{\bf Direct Neutralino Detection\ }  
In the mSUGRA model the $\chi^0_1-$nucleon scattering cross-section   
depends strongly on the values of $\tan\beta$ and $m_{\chi^0_1}$ \cite{jkg},
increasing with $\tan\beta$ but decreasing with $m_{\chi^0_1}$.   
The expected event rate in a $^{73}$Ge detector for   
$m_{\chi^0_1} \simeq 120$~GeV and $\tan\beta=35$ 
is about 0.015 events/kg-day~\cite{Brhlik}. 
The CDMS experiment in the Soudan mine expects to obtain 
a 2500 kg-day exposure on Germanium~\cite{cdms}.
The CDMS-Soudan experiment should observe at least 10 events 
due to neutralino scattering 
at $\tan\beta = 10$ with $m_{\chi^0_1} \alt 120$ GeV ($m_{1/2} \alt 300$ GeV) 
or 
at $\tan\beta = 35$ with $m_{\chi^0_1} \alt 160$ GeV ($m_{1/2} \alt 400$ GeV) 
\cite{Brhlik}.

\bigskip\noindent  
\underline{\bf Indirect Neutralino Detection\ }  
The CMB solutions at the largest $\tan\beta$ have neutralinos masses 
above 200~GeV. Large ice and water detectors  
(AMANDA, IceCube, ANTARES) can detect the high energy neutrinos from  
the annihilations of such heavy relic neutralinos that have  
gravitationally accumulated in the Sun~\cite{bhhk}.  
  
\bigskip\noindent  
\underline{\bf Intersection of Constraints\ }  
The $2 \sigma$ bands in Fig.~1 for $\Omega_{\rm cdm} h^2$ overlap 
the $2 \sigma$ ranges of the $\Delta a_\mu$ measurement, 
with different allowed $m_{1/2}$, $m_0$ mass regions in each $\tan\beta$ case.
If we ask for a more restrictive intersection of 
the $\Omega_{\rm cdm} h^2$ bands with the 1$\sigma$ allowed ranges 
for $\Delta a_\mu$, high $\tan\beta$ values of order 40 to 45 
are selected as shown in Fig.~2.

A further constraint of $m_h =$ 115 GeV gives $m_{1/2} \sim 300$ GeV 
and $m_0 \approx 350-550$ GeV for $\tan\beta \simeq 40-45$.
The corresponding SUSY particle masses are listed in Table II.

\bigskip\noindent  
\underline{\bf Infrared Fixed Point\ }  
In SUGRA models with approximate $b-\tau$ Yukawa unification 
at the grand unified scale, the renormalization group equations 
have solutions dominated by the quasi-fixed point of the top 
Yukawa coupling, which can be realized at high $\tan\beta$ \cite{BBOP,IRFP2}.
Moreover, high $\tan\beta$ texture models for the quark and the lepton
mass matrices \cite{DHR} require large values of $\tan\beta$.

\bigskip\noindent  
\underline{\bf Summary\ }  
We have determined the narrow regions of mSUGRA parameter space that can  
account for the restrictive cold dark matter density range 
inferred from recent CMB measurements. 
Our results are given in Figure~1 for four choices of $\tan\beta$.  
   
i) At $\tan\beta \alt 10$,  the $a_\mu$ measurement within 2$\sigma$ 
can be explained and a trilepton signal is predicted in Run\,II 
at the Tevatron; however, $m_h \alt 110$ GeV at $\tan\beta = 3$ makes 
this value of $\tan\beta$ less likely.  
   
ii) At $\tan\beta = 35$, both the measured $a_\mu$ within 2$\sigma$ 
and the $m_h = 115$~GeV hint from LEP\,2 are consistent with the CMB band;  
the trilepton signal here is marginal.  
   
iii) At $\tan\beta=50$, the allowed bands from 
$\Omega_{\rm cdm}h^2$ and $\Delta a_\mu$ 
are overlapping near the 2$\sigma$ limit of $\Delta a_\mu \sim 11$.  
  
iv) Consistency of $\Omega_{\rm cdm}h^2$ within the 1$\sigma$ uncertainty 
range of the $a_\mu$ measurement selects $\tan\beta \approx 40-45$.
We emphasize that this result does not depend on the Higgs mass.
The dependence of the solution on the trilinear coupling ($A_0$) 
is illustrated in Table I.

v) With $m_h = 115$~GeV as an additional constraint, supersymmetric particle 
masses are approximately determined as given in Table II.
The dependence on $\tan\beta$ is illustrated there. 
The predicted Higgs width is  about 5~MeV, 
compared to the SM Higgs width of 3~MeV.  
A broader Higgs width would ease beam resolution requirements for muon  
collider Higgs factories\cite{physrep}.  

Analysis of further data already taken on $a_\mu$~\cite{muon1} will reduce 
its uncertainty by about a factor of 2 and allow even closer specification 
of mSUGRA parameters derived from the overlap of $\Delta a_\mu$ 
and the CMB favored neutralino relic density.  
  
\section*{Acknowledgments}  
  
We are grateful to Michal Brhlik, Gi-Chol Cho, Toby Falk and Francis Halzen 
for beneficial discussions.  
This research was supported in part by the U.S. Department of Energy  
under Grants No. DE-FG02-95ER40896 and No. DE-FG03-98ER41066,   
and in part by the University of Wisconsin Research Committee  
with funds granted by the Wisconsin Alumni Research Foundation.  
  

\newpage  
  
\begin{table}  
\begin{tabular}{ccccccc}  
$A_0$ & $m_{\chi^0_1}$ & $m_{\chi^\pm_1}$ & $m_A$ & $m_h$ 
& $\Omega_{\chi^0_1} h^2$ & $\Delta a_\mu$ \\  
\hline  
$-300$ & 120 & 223 & 401 & 117 & 0.19 & 44 \\  
$-200$ & 120 & 221 & 398 & 116 & 0.16 & 45 \\  
$-100$ & 119 & 220 & 394 & 116 & 0.14 & 46 \\  
0      & 119 & 218 & 392 & 115 & 0.13 & 47 \\  
$+100$ & 119 & 217 & 388 & 115 & 0.11 & 48 \\  
$+200$ & 118 & 215 & 385 & 114 & 0.10 & 49 \\
$+300$ & 118 & 214 & 383 & 114 & 0.10 & 50 
\end{tabular}  
\caption{  
The relic density of neutralino dark matter ($\Omega_{\chi^0_1} h^2$),   
the SUSY contribution to the muon anomalous magnetic moment   
($\Delta a_\mu$) [Eq.~(\ref{eq:damuon})],  
and masses of the lightest neutralino ($\chi^0_1$),   
the lighter chargino ($\chi^\pm_1$),   
the CP-odd Higgs pseudoscalar ($A^0$),   
and the lighter CP-even Higgs scalar ($h^0$),   
for $\tan\beta = 40$, $m_{1/2} = 300$ GeV, $m_0 = 350$ GeV,   
and several values of the trilinear coupling ($A_0$).  
The masses and the trilinear coupling $A_0$ are in the units of GeV.  
}  
\label{table1}  
\end{table}  
  
\begin{table}  
\begin{tabular}{cccccccc}  
$\tan\beta$ & $m_A$ & $\Omega_{\chi^0_1} h^2$ & $\Delta a_\mu$ \\  
\hline  
35 & 453 &  0.32 & 37 \\  
37 & 437 &  0.25 & 39 \\  
40 & 411 &  0.16 & 42 \\  
42 & 391 &  0.12 & 44 \\  
45 & 360 &  0.06 & 48
\end{tabular}  
\caption{  
The relic density of neutralino dark matter ($\Omega_{\chi^0_1} h^2$),   
the SUSY contribution to the muon anomalous magnetic moment   
($\Delta a_\mu$) [Eq.~(\ref{eq:damuon})],  
and mass of the CP-odd Higgs pseudoscalar ($A^0$) in GeV,   
for $m_{1/2} = 300$ GeV, $m_0 = 400$ GeV, $A_0 = 0$,   
and several values of $\tan\beta$.  
For all these values of $\tan\beta$, 
$m_{\chi^0_1} =$ 119 GeV, $m_{\chi^\pm_1} =$ 218 GeV, $m_h =$ 115 GeV, 
and  $\Gamma_h = 4.8$ MeV, where $\Gamma_h$ is the decay width 
of the lighter CP-even Higgs boson ($h^0$).
}  
\label{table2}  
\end{table}  
  
\newpage  
  
\begin{figure}  
\centering\leavevmode  
\epsfxsize=6in  
\epsffile{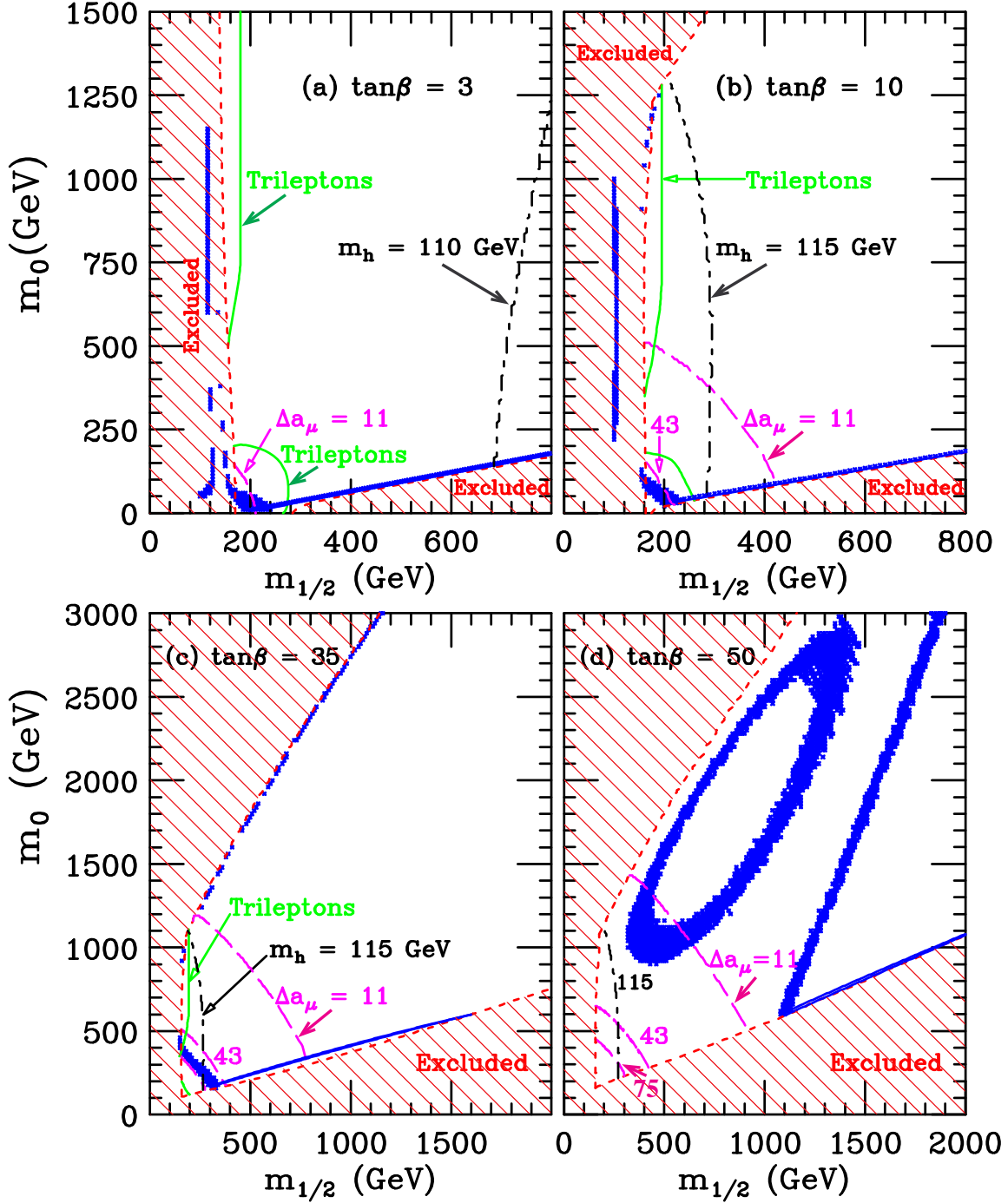}  
 
\bigskip 
\caption[]{  
Regions of relic density for neutralino dark matter satisfying   
the $2\sigma$ constraint from the new CMB data (dark blue bands)  
($0.11 \le \Omega_{\chi^0_1} h^2 \le 0.15$)  
in the ($m_{1/2},m_0$) plane of the mSUGRA model  
with $\mu >0$ and $A_0 = 0$ for (a) $\tan\beta = 3$,  
(b) $\tan\beta = 10$, (c) $\tan\beta=35$ and (d) $\tan\beta=50$.  
Also shown are the contours in the parameter space for   
(i) SUSY contributions of $\Delta a_\mu = 11$, 43, 75 
as defined in Eq.~(\ref{eq:damuon}), representing the measured value 
and its 2$\sigma$ uncertainties, 
(ii) the mass of the lighter CP even Higgs boson $m_h = 115$ GeV, and   
(iii) the 3$\sigma$ observation potential for a trilepton signal   
at the upgraded Fermilab Tevatron.  
The hatched regions are excluded by theoretical requirements   
[having electroweak symmetry breaking, the correct vacuum (tachyon free)   
and $\chi^0_1$ as the LSP]   
or by the chargino search at LEP\,2 \cite{SUSY}.  
\label{fig:omhs1}  
}\end{figure}  

\begin{figure}  
\centering\leavevmode  
\epsfxsize=6in  
\epsffile{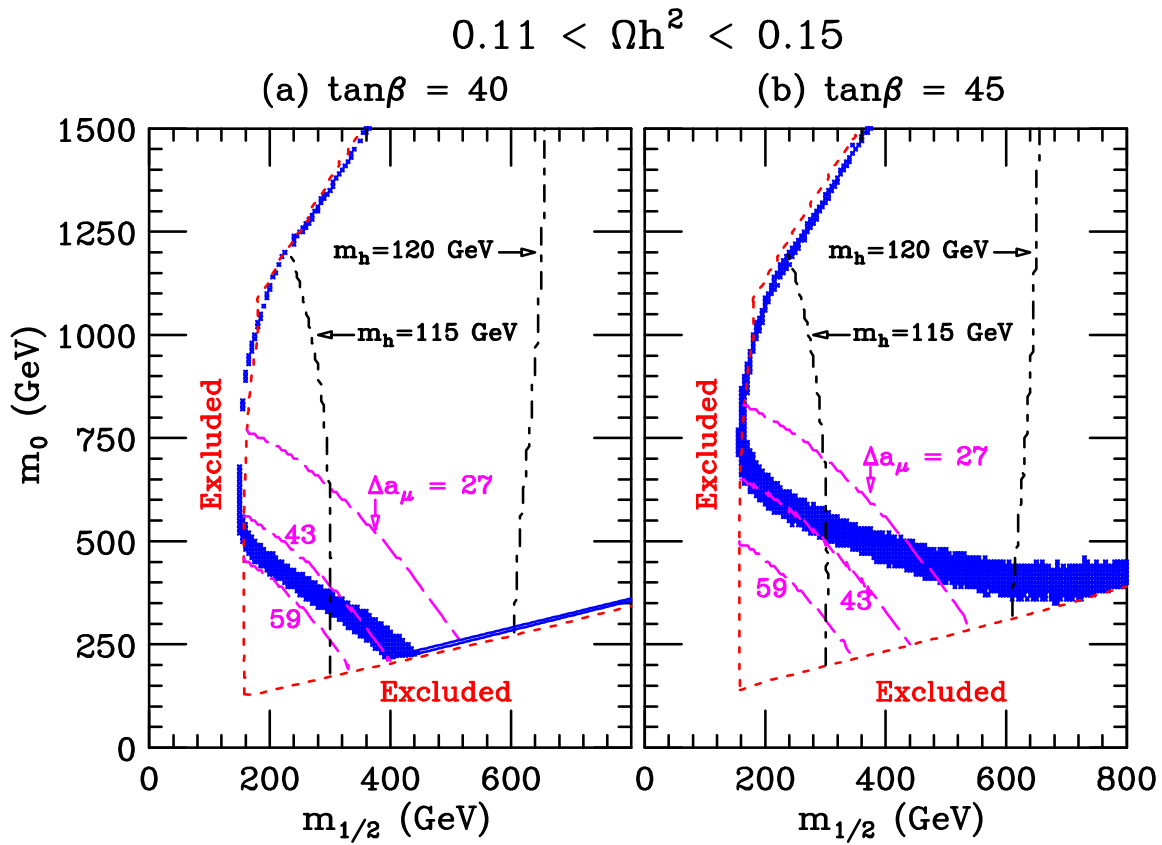}  
 
\bigskip 
\caption[]{  
Regions of the relic density allowed by the CMB data (dark blue bands)
for (a) $\tan\beta = 40$ and (b) $\tan\beta=45$; 
the SUSY contributions of $\Delta a_\mu = 27$, 43 and 59, 
represent the measured value and its 1$\sigma$ uncertainties. 
\label{fig:omhs2}  
}\end{figure}  
  
\end{document}